\documentclass[sigconf]{acmart}

\AtBeginDocument{%
  }

\copyrightyear{2023}
\acmYear{2023}
\setcopyright{acmlicensed}\acmConference[WWW '23 Companion]{Companion Proceedings of the ACM Web Conference 2023}{April 30-May 4, 2023}{Austin, TX, USA}
\acmBooktitle{Companion Proceedings of the ACM Web Conference 2023 (WWW '23 Companion), April 30-May 4, 2023, Austin, TX, USA}
\acmPrice{15.00}
\acmDOI{10.1145/3543873.3584624}
\acmISBN{978-1-4503-9419-2/23/04}

\usepackage{graphicx}
\usepackage{amsmath}
\usepackage{multirow}
\usepackage{booktabs}
\usepackage{array}
\usepackage{subfigure}
\usepackage{amssymb}
\usepackage{algorithm}
\usepackage{verbatim}
\usepackage{bm}     
\usepackage{enumitem}       

\usepackage{algpseudocode}

\begin{document}

\title{Reweighting Clicks with Dwell Time in Recommendation}


\author{Ruobing Xie}
\authornote{Both authors have equal contributions. Ruobing Xie is the corresponding author.}
\affiliation{\institution{WeChat, Tencent}
\city{Beijing}
\country{China}}
\email{ruobingxie@tencent.com}

\author{Lin Ma}
\authornotemark[1]
\affiliation{\institution{WeChat, Tencent}
\city{Beijing}
\country{China}}
\email{carrotma@tencent.com}

\author{Shaoliang Zhang}
\affiliation{\institution{WeChat, Tencent}
\city{Beijing}
\country{China}}
\email{modriczhang@tencent.com}

\author{Feng Xia}
\affiliation{\institution{WeChat, Tencent}
\city{Beijing}
\country{China}}
\email{xiafengxia@tencent.com}

\author{Leyu Lin}
\affiliation{\institution{WeChat, Tencent}
\city{Beijing}
\country{China}}
\email{goshawklin@tencent.com}

\renewcommand{\shortauthors}{Ruobing Xie et al.}

\begin{abstract}
The click behavior is the most widely-used user positive feedback in recommendation. However, simply considering each click equally in training may suffer from clickbaits and title-content mismatching, and thus fail to precisely capture users' real satisfaction on items. Dwell time could be viewed as a high-quality quantitative indicator of user preferences on each click, while existing recommendation models do not fully explore the modeling of dwell time. In this work, we focus on reweighting clicks with dwell time in recommendation. Precisely, we first define a new behavior named valid read, which helps to select high-quality click instances for different users and items via dwell time. Next, we propose a normalized dwell time function to reweight click signals in training for recommendation. The Click reweighting model achieves significant improvements on both offline and online evaluations in real-world systems.
\end{abstract}

\ccsdesc[500]{Information systems~Recommender systems}

\keywords{recommendation, valid read, dwell time, click reweighting}

\maketitle

\section{Introduction}

Real-world personalized recommendation attempts to provide appropriate items based on user preferences. User feedback on items is natural and essential information to discover user interests. \textbf{Click}, which is a high-quality and widely-existed implicit feedback, is the dominating user behavior used in recommendation. Click-through rate (CTR) prediction is also the central objective \cite{guo2017deepfm,zhou2018deep,xie2020deep}.

Despite the ubiquitous usage of clicks, simply relying on clicks as the only supervised training signals may not accurately and comprehensively capture users' real satisfaction, since the implicit click feedback often struggles with clickbaits or title-content mismatching in practice \cite{wang2021clicks}. Moreover, most existing recommendation models intuitively regard all clicks equally as training labels \cite{guo2017deepfm,sun2019bert4rec,liu2020autofis}, failing to dig out the different intensities of user preferences in each click.
To address these issues, an intuitive idea is to enhance the binary clicks with more quantified weights. \textbf{Dwell time (DT)}, which indicates the duration of a user on a clicked item (after clicking and before exiting), is easy-to-collect in real-world systems and perfectly suitable to quantify clicks and discover users' preferences \cite{yi2014beyond,kim2014modeling}. More dwell time indicates that users are more willing to pay time costs on items, reflecting higher user interests beyond clicks. Dwell time is also a widely-used online metric to measure users' real satisfaction in practical systems \cite{xie2022contrastive,xie2022long}.

\begin{figure}[!hbtp]
\centering
\includegraphics[width=0.92\columnwidth]{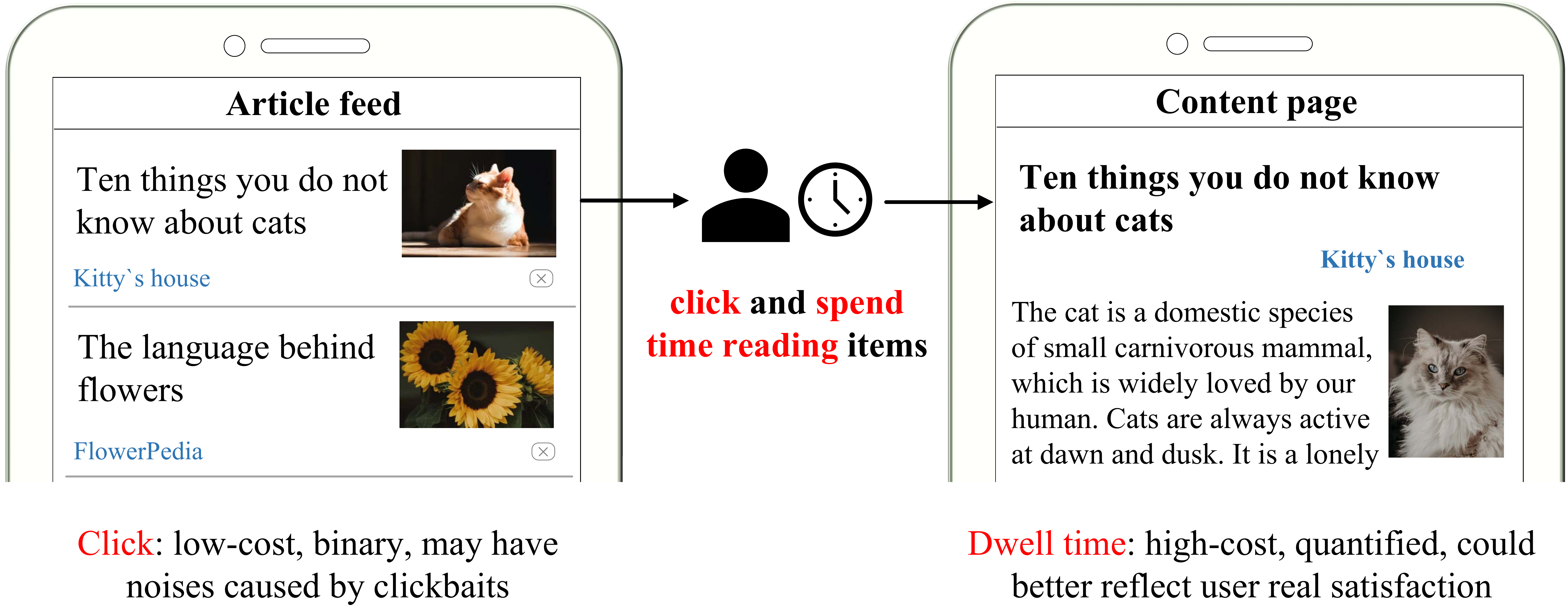}
\caption{Dwell time as the natural weights of clicks.}
\label{fig:example}
\end{figure}

There are a few works that jointly consider clicks and dwell time as objectives or features in practical recommendation \cite{zhou2018jump,chen2019follow,wu2020user,xie2021personalized}. However, most of them simply use the original/log dwell time as another training label besides clicks, ignoring further explorations on quantifying relations between dwell time and user satisfaction. Two questions need to be answered in dwell time modeling:

\textbf{\emph{(1) What a good recommender system should be?}}
We believe that a good recommender system should help users to get useful information more efficiently (rather than pursuing more clicks or dwell time). The central goal is to \emph{provide more valid readings} for users. Dwell time is intuitively used to define valid reading, while different users and items have different sensitivities on dwell time. For example, some users tend to spend less time reading (i.e., light users). An item's dwell time is also related to its type and total length (e.g., short news V.S. long videos). The valid readings of different users and items should be fairly considered in training.

\textbf{\emph{(2) How to accurately quantify user satisfaction with dwell time?}}
Longer dwell time does imply higher satisfaction, while the same dwell time improvement does not always indicate the same user satisfaction improvement. For example, the positive impact of a dwell time improvement from $1$s to $15$s is much larger than that from $601$s to $615$s. Intuitively, we hope users to have fewer invalid clicks with too-short dwell time, while we should also avoid over-emphasizing clicks with too-long dwell time, since the information gain will get lower and the seesaw effect may harm the learning of long-tail light users and short items. How to design a dwell time function to properly reweight clicks remains to be explored.

In this work, we aim to reweight clicks with dwell time to build a good recommender system, where users should \emph{have more high-quality and efficient readings}. Precisely, we propose a simple, effective, and model-agnostic \textbf{Click reweighting} framework to improve the training objectives.
First, we define a new behavior named \textbf{valid read} as a dwell time enhanced high-quality click behavior. The valid read selects three types of good clicks as the training signals, considering different demands of (a) the common-sense dwell time threshold learned from the global DT distribution, (b) light users, and (c) short items.
Second, we design a novel \textbf{normalized dwell time function} to quantify the posterior user satisfaction from dwell time on each valid read. We discover two characteristics a good normalized DT function should have to guild users to have more valid reads without much negative impact caused by behaviors having too-long dwell time. Finally, we conduct a multi-task learning (MTL) framework containing a valid read prediction tower and a reweighted valid read prediction tower.
As a first small step of click reweighting with dwell time, we intentionally select a rather straightforward industry-style model to enable facile expansions.

In experiments, we evaluate the Click reweighting framework on a real-world recommender system of WeChat Top Stories. Our model achieves significant improvements in both offline and online evaluations. The contributions of this work are as follows:
\begin{itemize}[leftmargin=*]
  \item We highlight the significance of valid read, rethink the quantification of user satisfaction with dwell time modeling, and propose our Click reweighting framework. To the best of our knowledge, we are the first to adopt the valid read behavior with dwell time based click reweighting in real-world recommender systems.
  \item We define the valid read to collect high-quality clicks considering the demands of different users and items. We also design a simple yet effective normalized dwell time function to model the intrinsic relationships between dwell time and user satisfaction.
  \item We evaluate our Click reweighting framework on both offline and online evaluations in a real-world recommender system, achieving significant improvements on various metrics. Currently, the proposed Click reweighting has been deployed on WeChat Top Stories for more than $4$ months, affecting millions of users.
\end{itemize}

\section{Model Designs and Analyses}

\subsection{Discussions on Dwell Time Modeling}
\label{sec.motivation}

Researchers have devoted themselves to exploring the core problem of recommendation: \emph{what kind of recommendation do users really need}. Recent efforts have shown the advantages of \emph{dwell time} in reflecting users' real satisfaction compared to \emph{CTR} \cite{zhou2018jump,chen2019follow,zheng2022dvr}.
However, directly optimizing raw dwell time will inevitably guide the model to over-emphasize items with long total durations, making heavy users and long items dominate the model training \cite{wang2020capturing,xie2021personalized}.

We believe that the central demand of users using recommender systems is to get information. Hence, we return to the essence of the relations among dwell time, information gain, and user preference, and conclude the following assumptions: (A1) the positive signals given by the same dwell time are relatively equivalent for different items and users, since they often imply the same valuable time cost that is fair to everyone.
(A2) Users need a minimum dwell time to begin to get information from items. Too-short dwell time implies very few (or no) benefits.
(A3) The information gain will gradually decrease with the increase of the dwell time when the current dwell time is long enough.
Based on these, we define the valid read with a normalized dwell time function in our click reweighting as a better supervised signal for more informative recommendations.

\subsection{Valid Read Selection}
\label{sec.valid_read}

\begin{figure}[!hbtp]
\centering
\includegraphics[width=0.99\columnwidth]{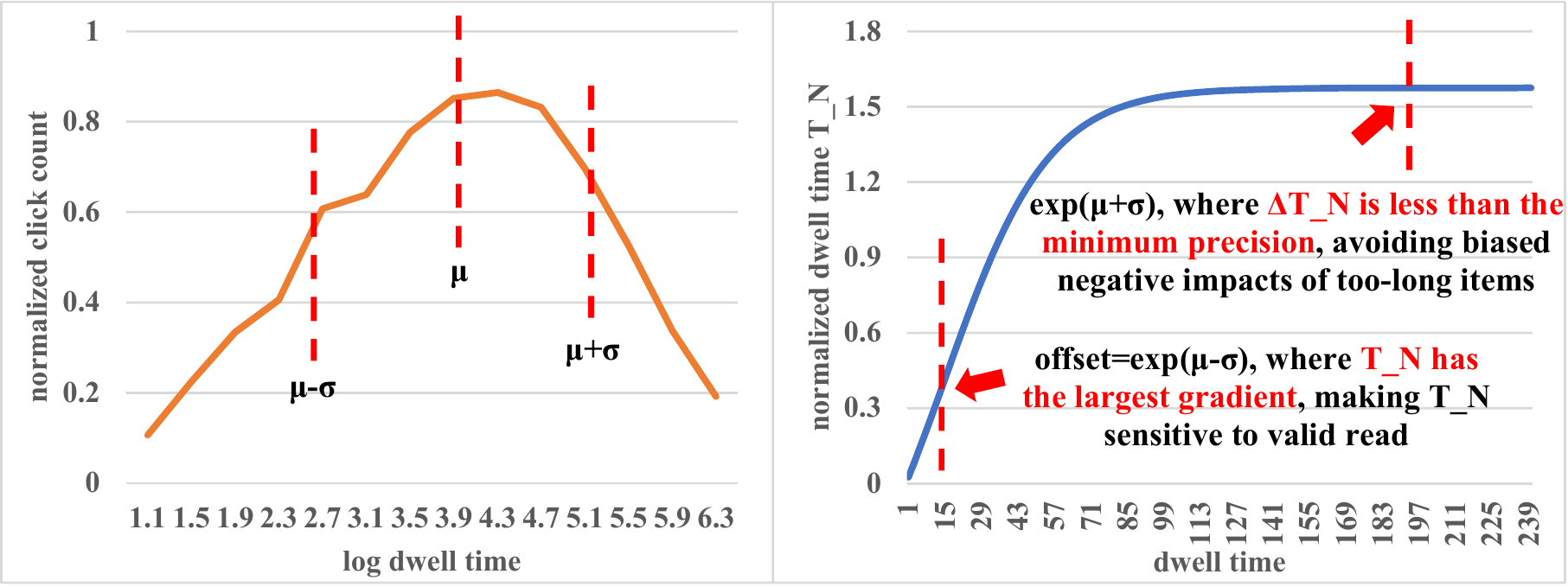}
\caption{The trend of log dwell time in our system (left) and the trend of our proposed normalized dwell time (right).}
\label{fig:DT_statistics}
\end{figure}

Valid reads are high-quality click behaviors that could better reflect users' real preferences, which are naturally selected via dwell time in this work. For a deeper understanding of dwell time, we draw the trends of click numbers with different log dwell time. From Fig. \ref{fig:DT_statistics} (left) we find that:
(1) in general, we could roughly assume that the log dwell time has an approximate Gaussian distribution, i.e., $\ln T=\mu+\sigma\epsilon$, where $T$ is a random dwell time and $\epsilon\sim N(0,1)$.
(2) We regard $[\mu-\sigma, \mu+\sigma]$ as the mainstream dwell time range. Nearly $19\%$ click behaviors have shorter than $15$s dwell time, and nearly $15\%$ click behaviors have longer than $200$s dwell time. According to the above assumptions A2 and A3, click behaviors with either too-short or too-long dwell time should be degraded in click reweighting.

It is straightforward to roughly set a shared dwell time threshold to collect valid reads. However, simply relying on the threshold to define valid reads will inevitably ignore the significant behavioral information of light users and short items.
Hence, we define three types of user-item clicks as our valid read behaviors:
\begin{itemize}
  \item T1: the dwell time is longer than $x_l$ seconds.
  \item T2: the user has clicked less than $7$ items in the recent week.
  \item T3: the dwell time is longer than $10\%$ of this item's historical dwell time records (i.e., longer than quantile P10).
\end{itemize}
(1) The first type builds the fundamental rule of valid read according to a common-sense threshold $x_l$. We regard $x_l=\exp(\mu-\sigma)$ of $\ln T$ as the shared dwell time threshold of valid read, which is adaptive to different recommender systems. In our system, $\exp(\mu-\sigma)$ is nearly $15$s. $19\%$ click behaviors are filtered by T1. For simplicity, we directly adopt a shared DT threshold for all users and items in respect of the absolute value of time costs, while it is also convenient to set customized dwell time thresholds for different user or item groups.
(2) The second type puts a patch on light users, considering all light users' click behaviors as supervised signals in training, for their behaviors are rare. We want to avoid critical information loss of long-tail light users that prefer scanning rather than deep reading.
(3) The third type considers the relative dwell time on a specific item, retrieving clicks that have a relatively qualified dwell time (top $90\%$) among all historical clicks on the same item. By this, our valid read shows respect to items with naturally short lengths and less dwell time (e.g., news or short videos).
To avoid noises, we further wipe out all clicks having less than $5$s dwell time to ensure the minimum availability of valid reads. In our practical system, the T1, T2, T3 types account for $89.9\%$, $2.9\%$, $7.2\%$ of the overall valid reads. Only valid reads are used as supervised signals in training.

\subsection{Normalized Dwell Time Function}
\label{sec.normalized_DT}

The valid read selection works as a pre-filter. However, we still face the challenge of precisely defining the goodness of different dwell time values in click reweighting. It is intuitive that the same dwell time improvement has a larger contribution to the quality of a click when the current dwell time is shorter (e.g., [1s$\rightarrow$15s] is larger than [601s$\rightarrow$615s]). Too long dwell time may bring in fatigue that harms user experience \cite{xie2022multi}. Hence, lots of works adopt MSE with log dwell time as training losses for dwell time prediction \cite{zhou2018jump,chen2019follow,wang2020capturing}.

Different from conventional models, we define the valid reads as our high-quality supervised labels and hope to improve the numbers and proportion of valid reads in online systems. Therefore, our dwell time function should possess the following two characteristics C1 and C2 respect to the above assumptions A2 and A3 in Sec. \ref{sec.motivation}:
\begin{itemize}
  \item C1: the designed dwell time function curve should be steep with large gradients in the early stage (especially near the valid read threshold $\exp(\mu-\sigma)$), guiding models to efficiently distinguish valid reads from invalid clicks.
  \item C2: the dwell time function curve should be flat when the dwell time is too long, avoiding too many rewards over long-duration items that harms light users and short items.
\end{itemize}
Following these rules, we design our \textbf{normalized dwell time} $T_N$ based on the original dwell time $T$ with a sigmoid function as:
\begin{equation}
\begin{split}
T_N=\frac{A}{1+\exp(-\frac{T-offset}{\tau})}-B.
\end{split}
\label{eq.L_semantic}
\end{equation}
Fig. \ref{fig:DT_statistics} (right) shows the trends of $T_N$.
$T_N$ is monotonically increasing with designed rates compared to log dwell time, where $offset$ and $\tau$ are essential parameters to satisfy C1 and C2.
$offset$ determines the dwell time point with the largest gradient. For C1, we set $offset=\exp(\mu-\sigma)$ to make the normalized dwell time have the largest gradient on the borderline of valid/invalid reads, which cooperates well with the valid read based supervised training. $\tau$ defines the sharpness of the dwell time curve.
For C2, we define an upper threshold $x_h$ as $\exp(\mu+\sigma)$, assuming that the dwell time $T$ larger than $x_h$ has no contribution on $T_N$ (i.e., the $T_N$ improvement of $x_h \rightarrow T$ is smaller than the minimum precision, e.g., $1e-5$ in our system). $\tau$ is set to fit the above assumption of $x_h$.
$A$ and $B$ are hyper-parameters that scale $T_N$ to $[0, T_{max}]$, where $T_{max}$ is the maximum dwell time value of our current online dwell time models. We remain the normalized dwell time range unchanged to reduce possible mismatching issues cooperating with other modules in online.
Finally, based on the above discussions, we set $offset=15$, $\tau=20$, $A=2.319$, $B=0.744$ to satisfy C1 and C2. We have also conducted a grid search on these parameters, and find that the current setting does achieve the best online performance.

\subsection{Click Reweighting}
\label{sec.optimization}

The valid read and normalized dwell time settings are designed to filter noises and quantify the qualities of clicks for better user preference learning. In Click reweighting, we adopt a multi-task learning (MTL) framework for both valid read prediction and weighted valid read prediction tasks. Specifically, we conduct a shared bottom to share raw user/item features across two tasks.

For the valid read tower,  without losing generality, we adopt a $3$-layer MLP, which takes raw user/item features $f_u, f_{d_i}$ as inputs, and outputs the predicted click probability $P_{u,d_i}$ of user $u$ on item $d_i$. Next, the valid read loss $L_v$ is defined as:
\begin{equation}
\begin{split}
L_v= - \sum_{(u,d_i) \in S_p} \log P_{u,d_i} + \sum_{(u,d_j) \in S_n} \log(1-P_{u,d_j}).
\end{split}
\label{eq.L_v}
\end{equation}
$S_p$ and $S_n$ indicate the positive (i.e., valid read) set and negative (i.e., invalid click and unclick) set, respectively. Similarly, for the weighted valid read tower, we directly use the normalized dwell time $T_N^{u,d_i}$ as the weight of each $(u,d_i)$. Another $3$-layer MLP is adopted to output the predicted click probability $P'_{u,d_i}$. The weighted valid read tower is then trained under the loss $L_w$ as follows:
\begin{equation}
\begin{split}
L_w=\sum_{(u,d_j) \in S_n} T_N^{u,d_j} \log(1-P'_{u,d_j}) - \sum_{(u,d_i) \in S_p} T_N^{u,d_i} \log P'_{u,d_i}.
\end{split}
\label{eq.L_w}
\end{equation}
$L_v$ and $L_w$ are linearly combined as the final loss $L=L_v+L_w$. In online deployment,
the sum of two towers' predicted scores is used for online ranking in our system.
Jointly considering the original and DT weighted valid read prediction tasks via MTL is beneficial for the overall online performance. Moreover, we have explored enhanced neural networks and MTL methods such as MMoE \cite{ma2018modeling} and PLE \cite{tang2020progressive}, while the online improvement is not significant. It may be because the dwell time is highly correlated with clicks. For simplicity, we directly use MLP with shared bottom in our model.

\section{Experiments}

\subsection{Dataset and Settings}

We conduct both offline and online evaluations on an article recommender system of WeChat Top Stories. The offline dataset contains nearly $29.7M$ users, $5.3M$ items, and $751M$ instances (including $104M$ clicks and $89.6M$ valid reads). All instances are chronologically split into the train set and the test set ($571M/180M$ instances). All data are pre-processed via data masking to protect user privacy.

\subsection{Offline Evaluation and Ablation Study}

We build four models with different objectives for offline evaluation and ablation study: (a) \textbf{Single CTR}, which only uses CTR as the training objective. (b) \textbf{CTR+logDT}, which is an MTL model with both CTR and log dwell time as objectives following classical CTR+DT optimization \cite{chen2019follow,wang2020capturing}. (c) \textbf{VR+logDT}, an MTL model with valid read (VR) and logDT objectives. (d) \textbf{VR+NDT} (i.e., the final Click reweighting model), which further replaces logDT with our normalized dwell time (NDT). We evaluate them on the valid read prediction task with AUC and RelaImpr as metrics following \cite{guo2017deepfm,song2019autoint}.
All baselines share the same neural network for single/MTL towers,  with the same raw features and settings for fair comparisons.

\begin{table}[!hbtp]
\centering
\small
\begin{tabular}{l|cccc}
\toprule
Model & single CTR & CTR+logDT & VR+logDT & VR+NDT \\
\midrule
AUC & 0.7810 & 0.7849 & 0.7932 & \textbf{0.7968} \\
RelaImpr & 0.00\% & 1.39\% & 4.34\% & \textbf{5.62\%} \\
\bottomrule
\end{tabular}
\caption{Offline evaluation on valid read prediction.}
\label{tab:offline}
\end{table}

Table \ref{tab:offline} shows the results, from which we find that:
(1) the final Click reweighting model achieves the best performance on valid read prediction. The improvements are significant ($p<0.01$ with paired t-test) and the deviations of all models are less than $\pm0.0003$. It indicates that our Click reweighting can recommend more high-quality items users love to click and read.
(2) Comparing models with/without VR and NDT, we find that both valid read filtering and normalized DT reweighting are essential for improving users' valid reads.
(3) Single CTR merely focuses on CTR, and thus performs worse than MTL models with dwell time modeling.
(4) We have also evaluated these models on the original CTR prediction, where single CTR achieves the best offline AUC for it is naturally designed for this task. However, the Click reweighting model surprisingly achieves the best CTR in the online A/B test (which we care more about). The click reweighting gives top priority to recommending \emph{high-quality items which users may have informative reads on}, rather than items users are likely to click guided by CTR-oriented training, bringing in long-term benefits via better user experience.

\subsection{Online Evaluation}

To verify the online power of Click reweighting, we further conduct an online A/B test on WeChat Top Stories. We focus on four online metrics: (a) CTR, (b) average click number per capita (ACN), (c) dwell time (DT), and (d) average impression number per capita (AIN). We conduct the A/B test for $7$ days on nearly $5$ million users.

\begin{table}[!hbtp]
\centering
\small
\begin{tabular}{l|cccc}
\toprule
Model & CTR & ACN & DT & AIN \\
\midrule
+valid read & +1.910\% & +2.478\% & -1.698\% & +0.558\% \\
+valid read+NDT & +2.577\% & +4.139\% & -3.554\% & +1.545\% \\
\bottomrule
\end{tabular}
\caption{Online A/B tests on WeChat Top Stories.}
\label{tab:online}
\end{table}

From Table \ref{tab:online} we find that:
(1) Both CTR and ACN have significant improvements ($p<0.05$) armed with valid read. It is impressive that using high-quality valid reads as training objectives can even improve the online click-related metrics. The improvements are further strengthened by adding normalized dwell time, which reconfirms the effectiveness of NDT on user experience.
(2) The original dwell time modeling over-emphasizes long dwell time behaviors. Our Click reweighting aims to improve valid reads for all users, thus inevitably sacrificing the performance of dwell time.
(3) The improvements in ACN and AIN further imply that users are more willing to use our system, which is the core driving force of growth.

\subsection{Online Dwell Time Migration}

\begin{figure}[!hbtp]
\centering
\includegraphics[width=0.90\columnwidth]{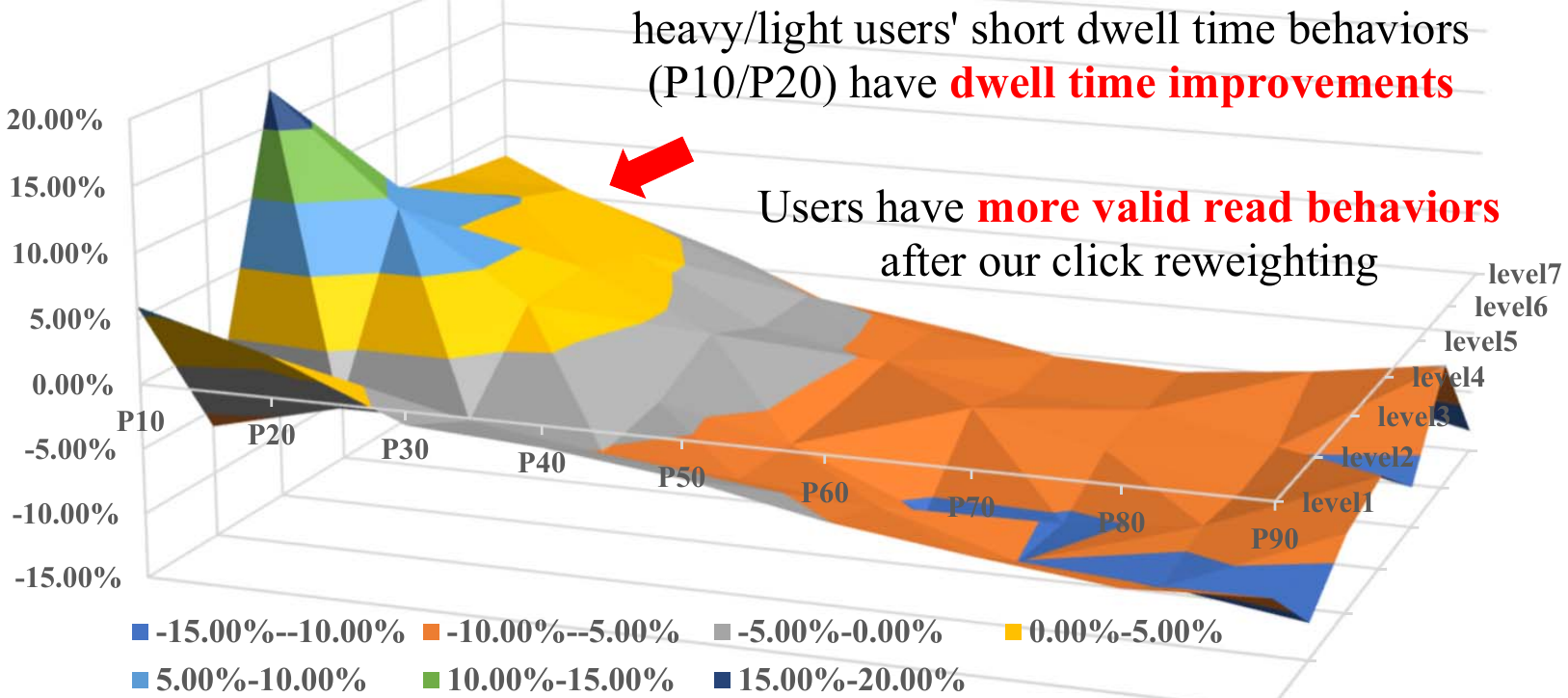}
\caption{Dwell time migration on different dwell time quantiles and different user activeness in an online system.}
\label{fig:DT_migration}
\end{figure}

In Fig. \ref{fig:DT_migration}, we discover an interesting dwell time migration trend of users with different activeness. The x-axis indicates the quantile of dwell time (P10 is the shortest $10\%$ dwell time), the y-axis indicates the activeness level (level7 is the most active users), and the z-axis indicates the dwell time changes from baseline to Click reweighting. We find that:
(1) both light and heavy users have more dwell time on their short dwell time behaviors (especially for less active users). It implies that users tend to have more valid reads.
(2) The dwell time of too-long readings inevitably decreases, since too-long items are not over-emphasized due to the normalized dwell time. In contrast, our model pays more attention to the behaviors of light users on short items.
(3) The DT migration matches our purpose to provide more informative and efficient recommendations. We hope users get a better reading experience rather than be stuck in our system.

\section{Related Works}

There are some efforts attempt to discover clickbaits and purify clicks \cite{chen2015misleading,potthast2016clickbait,agrawal2016clickbait,shang2019towards,kaur2020detecting}.
In real-world scenarios, the dwell time of clicked items is natural and powerful user feedback that can quantify clicks \cite{hassan2012semi,kim2014modeling,yi2014beyond,liu2016time, xie2022multi}. The content features are often carefully encoded for dwell time prediction \cite{wu2020user,wang2020capturing}. Recently, some works adopt MTL or multi-optimization objectives to jointly consider CTR and dwell time predictions \cite{zhou2018jump,chen2019follow,xie2021personalized}. However, they do not fully address the over-emphasizing issue of too-long items. \citet{zheng2022dvr} designs a watch time gain to measure the relative dwell time on an item, while it loses the essential information of the specific dwell time value in different items.
In this work, we propose a novel behavior valid read with a normalized DT to better fit our purpose of enabling more efficient and informative readings.

\section{Conclusion and Future Work}

In this work, we propose a simple yet effective way to reweight clicks via valid read based filtering with normalized dwell time based reweighting. The click reweighting framework has been deployed on a real-world recommender system in WeChat.
In the future, we will explore more sophisticated valid read modeling, and theoretically and experimentally investigate the pros and cons of our purposes of click reweighting via long-term online metrics.

\bibliographystyle{ACM-Reference-Format}
\bibliography{reference}


\begin{thebibliography}{27}


\ifx \showCODEN    \undefined \def \showCODEN     #1{\unskip}     \fi
\ifx \showDOI      \undefined \def \showDOI       #1{#1}\fi
\ifx \showISBNx    \undefined \def \showISBNx     #1{\unskip}     \fi
\ifx \showISBNxiii \undefined \def \showISBNxiii  #1{\unskip}     \fi
\ifx \showISSN     \undefined \def \showISSN      #1{\unskip}     \fi
\ifx \showLCCN     \undefined \def \showLCCN      #1{\unskip}     \fi
\ifx \shownote     \undefined \def \shownote      #1{#1}          \fi
\ifx \showarticletitle \undefined \def \showarticletitle #1{#1}   \fi
\ifx \showURL      \undefined \def \showURL       {\relax}        \fi
\providecommand\bibfield[2]{#2}
\providecommand\bibinfo[2]{#2}
\providecommand\natexlab[1]{#1}
\providecommand\showeprint[2][]{arXiv:#2}

\bibitem[Agrawal(2016)]%
        {agrawal2016clickbait}
\bibfield{author}{\bibinfo{person}{Amol Agrawal}.}
  \bibinfo{year}{2016}\natexlab{}.
\newblock \showarticletitle{Clickbait detection using deep learning}. In
  \bibinfo{booktitle}{\emph{Proceedings of NGCT}}.
\newblock


\bibitem[Chen et~al\mbox{.}(2019)]%
        {chen2019follow}
\bibfield{author}{\bibinfo{person}{Jingwu Chen}, \bibinfo{person}{Fuzhen
  Zhuang}, \bibinfo{person}{Tianxin Wang}, \bibinfo{person}{Leyu Lin},
  \bibinfo{person}{Feng Xia}, \bibinfo{person}{Lihuan Du}, {and}
  \bibinfo{person}{Qing He}.} \bibinfo{year}{2019}\natexlab{}.
\newblock \showarticletitle{Follow the Title Then Read the Article: Click-Guide
  Network for Dwell Time Prediction}.
\newblock \bibinfo{journal}{\emph{TKDE}} (\bibinfo{year}{2019}).
\newblock


\bibitem[Chen et~al\mbox{.}(2015)]%
        {chen2015misleading}
\bibfield{author}{\bibinfo{person}{Yimin Chen}, \bibinfo{person}{Niall~J
  Conroy}, {and} \bibinfo{person}{Victoria~L Rubin}.}
  \bibinfo{year}{2015}\natexlab{}.
\newblock \showarticletitle{Misleading online content: recognizing clickbait
  as" false news"}. In \bibinfo{booktitle}{\emph{Proceedings of the 2015 ACM on
  workshop on multimodal deception detection}}.
\newblock


\bibitem[Guo et~al\mbox{.}(2017)]%
        {guo2017deepfm}
\bibfield{author}{\bibinfo{person}{Huifeng Guo}, \bibinfo{person}{Ruiming
  Tang}, \bibinfo{person}{Yunming Ye}, \bibinfo{person}{Zhenguo Li}, {and}
  \bibinfo{person}{Xiuqiang He}.} \bibinfo{year}{2017}\natexlab{}.
\newblock \showarticletitle{DeepFM: a factorization-machine based neural
  network for CTR prediction}. In \bibinfo{booktitle}{\emph{Proceedings of
  IJCAI}}.
\newblock


\bibitem[Hassan(2012)]%
        {hassan2012semi}
\bibfield{author}{\bibinfo{person}{Ahmed Hassan}.}
  \bibinfo{year}{2012}\natexlab{}.
\newblock \showarticletitle{A semi-supervised approach to modeling web search
  satisfaction}. In \bibinfo{booktitle}{\emph{Proceedings of SIGIR}}.
\newblock


\bibitem[Kaur et~al\mbox{.}(2020)]%
        {kaur2020detecting}
\bibfield{author}{\bibinfo{person}{Sawinder Kaur}, \bibinfo{person}{Parteek
  Kumar}, {and} \bibinfo{person}{Ponnurangam Kumaraguru}.}
  \bibinfo{year}{2020}\natexlab{}.
\newblock \showarticletitle{Detecting clickbaits using two-phase hybrid
  CNN-LSTM biterm model}.
\newblock \bibinfo{journal}{\emph{Expert Systems with Applications}}
  (\bibinfo{year}{2020}).
\newblock


\bibitem[Kim et~al\mbox{.}(2014)]%
        {kim2014modeling}
\bibfield{author}{\bibinfo{person}{Youngho Kim}, \bibinfo{person}{Ahmed
  Hassan}, \bibinfo{person}{Ryen~W White}, {and} \bibinfo{person}{Imed
  Zitouni}.} \bibinfo{year}{2014}\natexlab{}.
\newblock \showarticletitle{Modeling dwell time to predict click-level
  satisfaction}. In \bibinfo{booktitle}{\emph{Proceedings of WSDM}}.
\newblock


\bibitem[Liu et~al\mbox{.}(2020)]%
        {liu2020autofis}
\bibfield{author}{\bibinfo{person}{Bin Liu}, \bibinfo{person}{Chenxu Zhu},
  \bibinfo{person}{Guilin Li}, \bibinfo{person}{Weinan Zhang},
  \bibinfo{person}{Jincai Lai}, \bibinfo{person}{Ruiming Tang},
  \bibinfo{person}{Xiuqiang He}, \bibinfo{person}{Zhenguo Li}, {and}
  \bibinfo{person}{Yong Yu}.} \bibinfo{year}{2020}\natexlab{}.
\newblock \showarticletitle{AutoFIS: Automatic Feature Interaction Selection in
  Factorization Models for Click-Through Rate Prediction}.
\newblock  (\bibinfo{year}{2020}).
\newblock


\bibitem[Liu et~al\mbox{.}(2016)]%
        {liu2016time}
\bibfield{author}{\bibinfo{person}{Yiqun Liu}, \bibinfo{person}{Xiaohui Xie},
  \bibinfo{person}{Chao Wang}, \bibinfo{person}{Jian-Yun Nie},
  \bibinfo{person}{Min Zhang}, {and} \bibinfo{person}{Shaoping Ma}.}
  \bibinfo{year}{2016}\natexlab{}.
\newblock \showarticletitle{Time-aware click model}.
\newblock \bibinfo{journal}{\emph{TOIS}} (\bibinfo{year}{2016}).
\newblock


\bibitem[Ma et~al\mbox{.}(2018)]%
        {ma2018modeling}
\bibfield{author}{\bibinfo{person}{Jiaqi Ma}, \bibinfo{person}{Zhe Zhao},
  \bibinfo{person}{Xinyang Yi}, \bibinfo{person}{Jilin Chen},
  \bibinfo{person}{Lichan Hong}, {and} \bibinfo{person}{Ed~H Chi}.}
  \bibinfo{year}{2018}\natexlab{}.
\newblock \showarticletitle{Modeling task relationships in multi-task learning
  with multi-gate mixture-of-experts}. In \bibinfo{booktitle}{\emph{Proceedings
  of KDD}}.
\newblock


\bibitem[Potthast et~al\mbox{.}(2016)]%
        {potthast2016clickbait}
\bibfield{author}{\bibinfo{person}{Martin Potthast}, \bibinfo{person}{Sebastian
  K{\"o}psel}, \bibinfo{person}{Benno Stein}, {and} \bibinfo{person}{Matthias
  Hagen}.} \bibinfo{year}{2016}\natexlab{}.
\newblock \showarticletitle{Clickbait detection}. In
  \bibinfo{booktitle}{\emph{Proceedings of ECIR}}.
\newblock


\bibitem[Shang et~al\mbox{.}(2019)]%
        {shang2019towards}
\bibfield{author}{\bibinfo{person}{Lanyu Shang}, \bibinfo{person}{Daniel~Yue
  Zhang}, \bibinfo{person}{Michael Wang}, \bibinfo{person}{Shuyue Lai}, {and}
  \bibinfo{person}{Dong Wang}.} \bibinfo{year}{2019}\natexlab{}.
\newblock \showarticletitle{Towards reliable online clickbait video detection:
  A content-agnostic approach}.
\newblock \bibinfo{journal}{\emph{Knowledge-Based Systems}}
  (\bibinfo{year}{2019}).
\newblock


\bibitem[Song et~al\mbox{.}(2019)]%
        {song2019autoint}
\bibfield{author}{\bibinfo{person}{Weiping Song}, \bibinfo{person}{Chence Shi},
  \bibinfo{person}{Zhiping Xiao}, \bibinfo{person}{Zhijian Duan},
  \bibinfo{person}{Yewen Xu}, \bibinfo{person}{Ming Zhang}, {and}
  \bibinfo{person}{Jian Tang}.} \bibinfo{year}{2019}\natexlab{}.
\newblock \showarticletitle{Autoint: Automatic feature interaction learning via
  self-attentive neural networks}. In \bibinfo{booktitle}{\emph{Proceedings of
  CIKM}}.
\newblock


\bibitem[Sun et~al\mbox{.}(2019)]%
        {sun2019bert4rec}
\bibfield{author}{\bibinfo{person}{Fei Sun}, \bibinfo{person}{Jun Liu},
  \bibinfo{person}{Jian Wu}, \bibinfo{person}{Changhua Pei},
  \bibinfo{person}{Xiao Lin}, \bibinfo{person}{Wenwu Ou}, {and}
  \bibinfo{person}{Peng Jiang}.} \bibinfo{year}{2019}\natexlab{}.
\newblock \showarticletitle{BERT4Rec: Sequential Recommendation with
  Bidirectional Encoder Representations from Transformer}. In
  \bibinfo{booktitle}{\emph{Proceedings of CIKM}}.
\newblock


\bibitem[Tang et~al\mbox{.}(2020)]%
        {tang2020progressive}
\bibfield{author}{\bibinfo{person}{Hongyan Tang}, \bibinfo{person}{Junning
  Liu}, \bibinfo{person}{Ming Zhao}, {and} \bibinfo{person}{Xudong Gong}.}
  \bibinfo{year}{2020}\natexlab{}.
\newblock \showarticletitle{Progressive layered extraction (ple): A novel
  multi-task learning (mtl) model for personalized recommendations}. In
  \bibinfo{booktitle}{\emph{Proceedings of RecSys}}.
\newblock


\bibitem[Wang et~al\mbox{.}(2020)]%
        {wang2020capturing}
\bibfield{author}{\bibinfo{person}{Tianxin Wang}, \bibinfo{person}{Jingwu
  Chen}, \bibinfo{person}{Fuzhen Zhuang}, \bibinfo{person}{Leyu Lin},
  \bibinfo{person}{Feng Xia}, \bibinfo{person}{Lihuan Du}, {and}
  \bibinfo{person}{Qing He}.} \bibinfo{year}{2020}\natexlab{}.
\newblock \showarticletitle{Capturing Attraction Distribution: Sequential
  Attentive Network for Dwell Time Prediction}. In
  \bibinfo{booktitle}{\emph{Proceedings of 2020}}.
\newblock


\bibitem[Wang et~al\mbox{.}(2021)]%
        {wang2021clicks}
\bibfield{author}{\bibinfo{person}{Wenjie Wang}, \bibinfo{person}{Fuli Feng},
  \bibinfo{person}{Xiangnan He}, \bibinfo{person}{Hanwang Zhang}, {and}
  \bibinfo{person}{Tat-Seng Chua}.} \bibinfo{year}{2021}\natexlab{}.
\newblock \showarticletitle{Clicks can be cheating: Counterfactual
  recommendation for mitigating clickbait issue}. In
  \bibinfo{booktitle}{\emph{Proceedings of SIGIR}}.
\newblock


\bibitem[Wu et~al\mbox{.}(2020)]%
        {wu2020user}
\bibfield{author}{\bibinfo{person}{Chuhan Wu}, \bibinfo{person}{Fangzhao Wu},
  \bibinfo{person}{Tao Qi}, {and} \bibinfo{person}{Yongfeng Huang}.}
  \bibinfo{year}{2020}\natexlab{}.
\newblock \showarticletitle{User Modeling with Click Preference and Reading
  Satisfaction for News Recommendation.}. In
  \bibinfo{booktitle}{\emph{Proceedings of IJCAI}}.
\newblock


\bibitem[Xie et~al\mbox{.}(2020)]%
        {xie2020deep}
\bibfield{author}{\bibinfo{person}{Ruobing Xie}, \bibinfo{person}{Cheng Ling},
  \bibinfo{person}{Yalong Wang}, \bibinfo{person}{Rui Wang},
  \bibinfo{person}{Feng Xia}, {and} \bibinfo{person}{Leyu Lin}.}
  \bibinfo{year}{2020}\natexlab{}.
\newblock \showarticletitle{Deep feedback network for recommendation}. In
  \bibinfo{booktitle}{\emph{Proceedings of IJCAI}}.
\newblock


\bibitem[Xie et~al\mbox{.}(2022a)]%
        {xie2022multi}
\bibfield{author}{\bibinfo{person}{Ruobing Xie}, \bibinfo{person}{Cheng Ling},
  \bibinfo{person}{Shaoliang Zhang}, \bibinfo{person}{Feng Xia}, {and}
  \bibinfo{person}{Leyu Lin}.} \bibinfo{year}{2022}\natexlab{a}.
\newblock \showarticletitle{Multi-granularity Fatigue in Recommendation}. In
  \bibinfo{booktitle}{\emph{Proceedings of CIKM}}.
\newblock


\bibitem[Xie et~al\mbox{.}(2022b)]%
        {xie2022contrastive}
\bibfield{author}{\bibinfo{person}{Ruobing Xie}, \bibinfo{person}{Qi Liu},
  \bibinfo{person}{Liangdong Wang}, \bibinfo{person}{Shukai Liu},
  \bibinfo{person}{Bo Zhang}, {and} \bibinfo{person}{Leyu Lin}.}
  \bibinfo{year}{2022}\natexlab{b}.
\newblock \showarticletitle{Contrastive cross-domain recommendation in
  matching}. In \bibinfo{booktitle}{\emph{Proceedings of KDD}}.
\newblock


\bibitem[Xie et~al\mbox{.}(2021)]%
        {xie2021personalized}
\bibfield{author}{\bibinfo{person}{Ruobing Xie}, \bibinfo{person}{Yanlei Liu},
  \bibinfo{person}{Shaoliang Zhang}, \bibinfo{person}{Rui Wang},
  \bibinfo{person}{Feng Xia}, {and} \bibinfo{person}{Leyu Lin}.}
  \bibinfo{year}{2021}\natexlab{}.
\newblock \showarticletitle{Personalized approximate pareto-efficient
  recommendation}. In \bibinfo{booktitle}{\emph{Proceedings of WWW}}.
\newblock


\bibitem[Xie et~al\mbox{.}(2022c)]%
        {xie2022long}
\bibfield{author}{\bibinfo{person}{Ruobing Xie}, \bibinfo{person}{Yalong Wang},
  \bibinfo{person}{Rui Wang}, \bibinfo{person}{Yuanfu Lu},
  \bibinfo{person}{Yuanhang Zou}, \bibinfo{person}{Feng Xia}, {and}
  \bibinfo{person}{Leyu Lin}.} \bibinfo{year}{2022}\natexlab{c}.
\newblock \showarticletitle{Long short-term temporal meta-learning in online
  recommendation}. In \bibinfo{booktitle}{\emph{Proceedings of WSDM}}.
\newblock


\bibitem[Yi et~al\mbox{.}(2014)]%
        {yi2014beyond}
\bibfield{author}{\bibinfo{person}{Xing Yi}, \bibinfo{person}{Liangjie Hong},
  \bibinfo{person}{Erheng Zhong}, \bibinfo{person}{Nanthan~Nan Liu}, {and}
  \bibinfo{person}{Suju Rajan}.} \bibinfo{year}{2014}\natexlab{}.
\newblock \showarticletitle{Beyond clicks: dwell time for personalization}. In
  \bibinfo{booktitle}{\emph{Proceedings of RecSys}}.
\newblock


\bibitem[Zheng et~al\mbox{.}(2022)]%
        {zheng2022dvr}
\bibfield{author}{\bibinfo{person}{Yu Zheng}, \bibinfo{person}{Chen Gao},
  \bibinfo{person}{Jingtao Ding}, \bibinfo{person}{Lingling Yi},
  \bibinfo{person}{Depeng Jin}, \bibinfo{person}{Yong Li}, {and}
  \bibinfo{person}{Meng Wang}.} \bibinfo{year}{2022}\natexlab{}.
\newblock \showarticletitle{DVR: Micro-Video Recommendation Optimizing
  Watch-Time-Gain under Duration Bias}. In \bibinfo{booktitle}{\emph{MM}}.
\newblock


\bibitem[Zhou et~al\mbox{.}(2018b)]%
        {zhou2018deep}
\bibfield{author}{\bibinfo{person}{Guorui Zhou}, \bibinfo{person}{Xiaoqiang
  Zhu}, \bibinfo{person}{Chenru Song}, \bibinfo{person}{Ying Fan},
  \bibinfo{person}{Han Zhu}, \bibinfo{person}{Xiao Ma},
  \bibinfo{person}{Yanghui Yan}, \bibinfo{person}{Junqi Jin},
  \bibinfo{person}{Han Li}, {and} \bibinfo{person}{Kun Gai}.}
  \bibinfo{year}{2018}\natexlab{b}.
\newblock \showarticletitle{Deep interest network for click-through rate
  prediction}. In \bibinfo{booktitle}{\emph{Proceedings of KDD}}.
\newblock


\bibitem[Zhou et~al\mbox{.}(2018a)]%
        {zhou2018jump}
\bibfield{author}{\bibinfo{person}{Tengfei Zhou}, \bibinfo{person}{Hui Qian},
  \bibinfo{person}{Zebang Shen}, \bibinfo{person}{Chao Zhang},
  \bibinfo{person}{Chengwei Wang}, \bibinfo{person}{Shichen Liu}, {and}
  \bibinfo{person}{Wenwu Ou}.} \bibinfo{year}{2018}\natexlab{a}.
\newblock \showarticletitle{Jump: A joint predictor for user click and dwell
  time}. In \bibinfo{booktitle}{\emph{Proceedings of IJCAI}}.
\newblock


\end{thebibliography}

\end{document}